\documentclass[final,1p,authoryear]{elsarticle}
\usepackage[T1]{fontenc}
\usepackage[utf8]{inputenc}
\usepackage{lmodern}
\usepackage{amsmath,amssymb,booktabs,microtype,graphicx}
\usepackage{geometry}
\usepackage{float}
\usepackage{xcolor}
\usepackage[colorlinks=true,linkcolor=blue,citecolor=blue,urlcolor=blue]{hyperref}
\newcommand{\clip}{\operatorname{clip}}
\newcommand{\sgn}{\operatorname{sgn}}

\begin{document}
\begin{frontmatter}
\title{An alternative to origin–destination matrix adjustment: Calibrating mode choice from traffic counts in MATSim}
\author[csfm]{Abdelkader Dib}
\author[csfm]{Milos Bala{\'c}}
\affiliation[csfm]{organization={Center for Sustainable Future Mobility (CSFM), ETH Zurich},city={Zurich},country={Switzerland}}
\begin{abstract}
Traffic counts are often used to calibrate transport models by adjusting origin--destination demand. We present an alternative for activity-based models that adjusts agent-specific car alternative-specific constants while retaining activity locations. Differences between observed and simulated counts guide utility corrections during successive MATSim iterations, while aggregate constants regulate mode shares. We implement the method in eqasim and test it in the Swiss national MATSim scenario. After 400 iterations, the share of 2,371 count observations with GEH at most three increases from 50.7\% to 66.5\%, and the share with GEH at most five increases from 73.5\% to 84.4\%. The calibrated constants show heterogeneous corrections to car propensity. The approach offers a way to improve the spatial allocation of car use when the underlying activity demand is representative.
\end{abstract}
\begin{keyword}
Activity-based modelling \sep Mode choice \sep Demand calibration \sep Alternative-specific constants \sep Traffic counts \sep MATSim
\end{keyword}
\end{frontmatter}

\section{Introduction}
Traffic counts record how much traffic passes a location. They reveal little about where those trips began, where they end or who made them. In four-step models, a common response to a count discrepancy is to adjust the origin--destination (OD) matrix. Information-based methods and generalised least squares reconcile a prior matrix with flows observed on the network \citep{vanzuylen1980,cascetta1984}. More recent methods estimate OD demand and path choice jointly \citep{cao2024}. In each case, the interpretation of a mismatch depends on what the analyst allows to change.

In an activity-based model, an OD pair connects two activities in a person's day. Synthetic population methods use census and travel survey data to assign homes, workplaces and daily schedules \citep{hoerl2021pipeline,hoerl2021population}. These assignments carry information that would be lost if destinations were moved solely to match traffic counts. Where the number and location of activities have independent support, a useful first question is whether the right trips are being made by car. This distinction matters for validation because several combinations of demand and assignment can fit the same observations \citep{cascetta1984,bastarianto2023}.

In MATSim, Cadyts uses counts to update the probabilities of individual travel plans \citep{flotterod2011}. Its standard integration adds a correction to the score of each plan, influencing which plan an agent selects in subsequent iterations \citep{horni2016,nagel2016,hoerl2021scoring}. This can serve as a disaggregate form of OD calibration when the available plans contain different destinations. For example, \citet{ziemke2015} used Berlin traffic counts to calibrate the selection of activity plans with different locations. If the plans share their trip ends, Cadyts can still adjust other choices represented among them. The eqasim pipeline generates a synthetic population and daily activity schedules from sources such as census records and travel surveys, then converts them into MATSim inputs \citep{hoerl2021pipeline}. Its default mode-choice approach uses the MATSim discrete mode choice extension \citep{hoerl2018,hoerl2021scoring}. Each agent keeps a single daily plan. At each iteration, a subset of agents re-evaluates the feasible modes for its trips using the choice model and current network conditions. This decision takes place without scoring and selecting among multiple stored plans. We introduce the count correction directly into the utility used for that decision.

An alternative-specific constant (ASC) captures the average effect of factors that the utility function does not model explicitly, relative to a reference mode \citep{train2009,rakow2025}. Travellers' preferences for comfort and flexibility can influence mode choice \citep{johansson2006}. Perceived safety can also affect the appeal of an e-bike for a trip to or from a public transport stop \citep{tzouras2025}. Preferences can vary with local accessibility \citep{orrego2023}, and empirical models represent some of this variation by estimating distributions of constants across travellers \citep{capurso2021,reck2022}. A distribution alone leaves open a spatial question. If four of ten travellers drive, aggregate car use is the same whichever four drive. Traffic on a particular road depends on their origins and destinations.

We propose calibrating individual car ASCs as an alternative to adjusting OD demand directly from traffic counts. Aggregate mode constants have already been calibrated automatically in the Swiss national scenario \citep{dib2026online,dib2026mum} and in Open Berlin \citep{rakow2025}. We retain aggregate constants to regulate mode shares and introduce one car constant per eligible traveller. This allows the car OD matrix to change as travellers switch modes while preserving the underlying trips across all modes. We examine count agreement and the evolution of the individual constants over 400 iterations of the Swiss national model.

\section{Methodology}
A common approach to simulation-based calibration is to run the model for one set of parameters, evaluate its outputs and run it again for the next set. Following \citet{dib2026online,dib2026mum}, we update the parameters during a single MATSim run. At selected iteration ends, we compare simulated traffic with counts, associate the discrepancies with OD pairs and average their contributions over each agent's daily trips. As agents subsequently reconsider their modes, the new choices affect traffic in later iterations.

\subsection{Agent and aggregate utility corrections}
Let $n$, $j$, $m$ and $t$ index agents, trips, modes and simulation iterations. The systematic utility is
\begin{equation}
 V_{njm}^{(t)}=V_{njm}^{0,(t)}+\alpha_{g(n),m}^{(t)}
                  +\mathbb I(m=\mathrm{car})a_n^{(t)}.
 \label{eq:utility}
\end{equation}
The base utility $V^0$ includes measured attributes and the original mode constants. The offset $\alpha_{g(n),m}$ responds to the mode share in agent $n$'s group, which may be a region or the whole country. The individual constant $a_n$ enters the car utility for every trip by that agent. A positive increment makes driving more attractive wherever car is a feasible alternative. Activity locations, the set of available modes and the model's trip and tour constraints are preserved.

\subsection{Count discrepancies and spatial pooling}
We assign each trip to the ordered pair $q$ of zones containing its origin and destination. Executed car routes identify which counted links serve that pair. We pool the errors at those links to obtain one signed correction for the pair. A traveller who currently takes public transport can therefore receive a car-utility correction if drivers travelling between the same zones encounter a count discrepancy.

For counted link $l$, let $C_l$ and $F_l^{(t)}$ denote observed and simulated flow expressed on the same scale. The station signal and its pooled OD correction are
\begin{align}
 e_l^{(t)}&=\clip\left(\log\frac{C_l+\epsilon}{F_l^{(t)}+\epsilon},-E,E\right),
 \label{eq:error}\\
 G_q^{(t)}&=\frac{N_q^{(t)}}{N_q^{(t)}+\kappa s}
 \frac{\sum_{(j,l)\in\mathcal O_q^{(t)}}w_{jl}e_l^{(t)}}
      {\sum_{(j,l)\in\mathcal O_q^{(t)}}w_{jl}},
 \qquad w_{jl}=\max\{K_j^{-1},w_{\min}\}\,h(C_l)\rho_l.
 \label{eq:pool}
\end{align}
The set $\mathcal O_q^{(t)}$ contains the accepted passages of car trips in group $q$ at counted links. Its size $N_q^{(t)}$ measures the amount of available evidence. The first factor in Eq.~\eqref{eq:pool} reduces the influence of sparsely observed OD pairs, with shrinkage parameter $\kappa$ and population sampling fraction $s$. Each passage is weighted by the number $K_j$ of counted links on the trip, the count-volume weight $h(C_l)$ and the station weight $\rho_l$. The floor $w_{\min}$ limits the reduction on trips that pass many stations. One trip can thus contribute several correlated observations.

A positive $e_l^{(t)}$ indicates that the simulated flow is too low. We exclude links with nonpositive observed or simulated flow and links within the relative tolerance. The constant $\epsilon$ stabilises the ratio and $E$ bounds its logarithm. An OD pair without an accepted passage receives a correction of zero. Consequently, a station with observed traffic but no simulated passages cannot by itself supply a correction.

\subsection{Updating agents and refining the zones}
For agent $n$, let $\mathcal J_n$ contain all trips with valid mapped OD pairs. At each scheduled update,
\begin{align}
 d_n^{(t)}&=\frac{\eta_t}{|\mathcal J_n|}\sum_{j\in\mathcal J_n}G_{q(j)}^{(t)},
 \qquad \eta_t=\max\{\eta_{\min},\eta_0\gamma^{\max(0,t-t_0)}\},
 \label{eq:delta}\\
 a_n^{(t+)}&=\clip\big(a_n^{(t)}+B(d_n^{(t)}),-A,A\big),
 \quad B(d)=
 \begin{cases}0,& |d|<b,\\
 \sgn(d)\min\{d_{\max},\max(d_{\min},|d|)\},&\text{otherwise}.
 \end{cases}
 \label{eq:update}
\end{align}
We use one daily correction per agent to keep the method simple. Each valid trip contributes once to the average, including trips whose OD correction is zero. An undercount on a morning commute and an overcount on an evening shopping trip can therefore offset one another. Hourly counts would allow separate corrections by time of day. After warmup $t_0$, the learning rate declines to a lower bound. The threshold $b$ suppresses small daily corrections, and $d_{\min}$ and $d_{\max}$ bound the size of an accepted step. For efficiency, the calculation uses the car routes already travelled in the simulation, so calibration requires no additional routing.

We start with large square zones to capture broad patterns, such as traffic on motorways that serve several towns. The grid divides a square into four smaller squares when it contains more eligible residents than the population threshold and its width is above the size limit. Dense urban areas therefore have finer zones than sparsely populated areas. At selected iterations we rebuild the grid with smaller size limits and lower population thresholds. Two trips that initially shared the same pair of zones can then receive different corrections if they serve different local corridors. Agents carry their existing constants into each finer grid.

\subsection{Aggregate mode-share control}
For each group $g$ and mode $m$, we compare the target share $z_{gm}$ with the share $\hat p_{gm}^{(t)}$ estimated from the agents replanning at iteration $t$. At a scheduled update, we add a fraction $\lambda_g^{(t)}$ of the log share discrepancy, measured relative to reference mode $r$, to the current constant. The next iteration uses the updated value given by
\begin{equation}
 \alpha_{gm}^{(t+1)}=\alpha_{gm}^{(t)}+\lambda_g^{(t)}
 \left[\log\frac{z_{gm}}{\hat p_{gm}^{(t)}}-\log\frac{z_{gr}}{\hat p_{gr}^{(t)}}\right],
 \qquad \alpha_{gr}^{(t+1)}=0.
 \label{eq:aggregate}
\end{equation}
Between updates, the constants carry forward. A small positive floor on the shares keeps the logarithms finite. The target shares anchor the overall frequency of car use. The individual constants allow the spatial allocation of that use to change. The Swiss application uses public transport as the reference mode and applies the controller at national and canton levels. A common shift in the individual car constants can be offset by the aggregate car constant, so their separate magnitudes have no unique behavioural interpretation.

\section{Swiss national application}
\subsection{Experimental design}
We use a 10\% population sample from the Swiss national eqasim scenario \citep{dib2026swiss}. In both runs, aggregate ASCs are calibrated for the cantons, Switzerland's constituent regions. Both runs also calibrate road-category penalties in the routing cost and network free-speed factors. We compare traffic flows at iteration 400 in runs with and without agent-specific car ASC calibration.

All individual ASCs start at zero. Following a 20-iteration warmup, updates take place every 15 iterations from iteration 30 to 390. We set $\eta_0=4$, $\eta_{\min}=0.3$, $\gamma=0.98$ and $A=3$. The daily update threshold is 0.005, and accepted steps range from 0.03 to 1.0. For the station calculation we use a 2\% flow tolerance, $\epsilon=1$, $E=1.5$, $\kappa=100$ and $w_{\min}=0.08$. For the volume weight $h(C_l)$ in Eq.~\eqref{eq:pool}, we rank counted links by observed flow and divide them into four groups at the quartiles. Links in the two lower quartiles receive a weight of 1.0 and those in the two upper quartiles a weight of 1.2. Other weights being equal, a passage at a higher-volume link therefore contributes 20\% more to the OD correction.

The initial grid starts with squares 16~km wide. A square wider than 2~km is split when it contains more than 200 eligible residents in the simulated sample, equivalent to a full-population threshold of 2,000. At iterations 75, 120 and 165, the starting widths are 10, 8 and 8~km, the minimum-width settings are 800, 500 and 200~m, and the corresponding thresholds in the sample are 100, 80 and 50 residents. The number of recorded home zones grows from 4,410 to 37,052.

We apply Eq.~\eqref{eq:aggregate} first at the national level (iterations 2--8). The national target shares are 41.48\% for car drivers, 15.20\% for public transport, 25.79\% for walking, 8.31\% for cycling and 9.22\% for car passengers. At iterations 9--99, the same equation is applied separately in each canton using its own target shares. The resulting canton offsets are retained for the remainder of the run.

\subsection{Count agreement}
We compare weekday traffic against 2{,}371 observations from Swiss federal and cantonal sources. We additionally report a separate comparison for 216 observations from the Swiss Federal Roads Office (ASTRA), which are located primarily on major roads. The evaluation uses the flows of iteration~400, expanded to the full population.

\begin{figure}[!htbp]\centering
\begin{minipage}{0.8\linewidth}\centering
    \begin{minipage}{0.47\linewidth}\centering
    (a) Without agent calibration\\[2pt]
    \includegraphics[width=\linewidth]{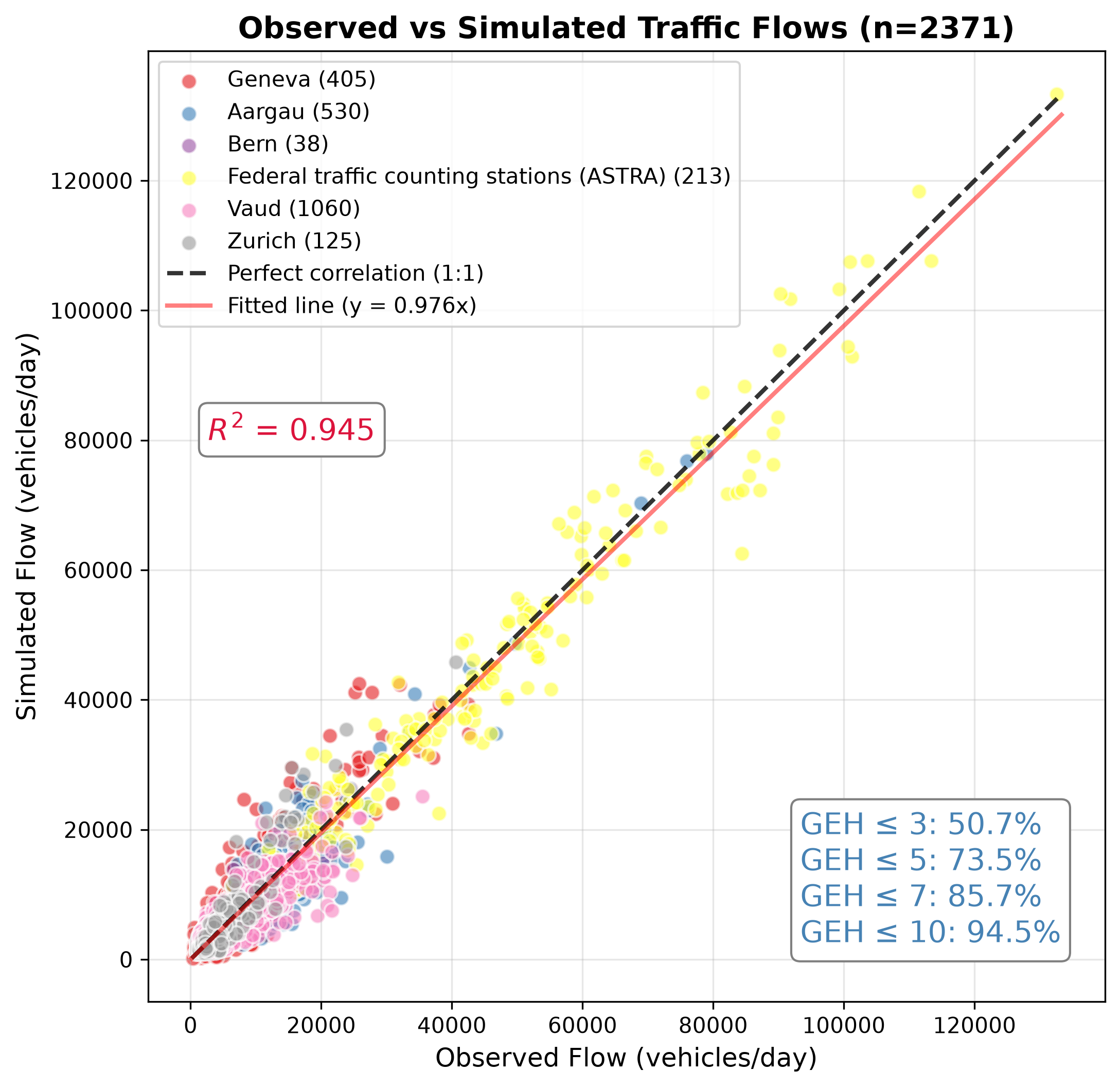}
    \end{minipage}\hfill
    \begin{minipage}{0.47\linewidth}\centering
    (b) With agent calibration\\[2pt]
    \includegraphics[width=\linewidth]{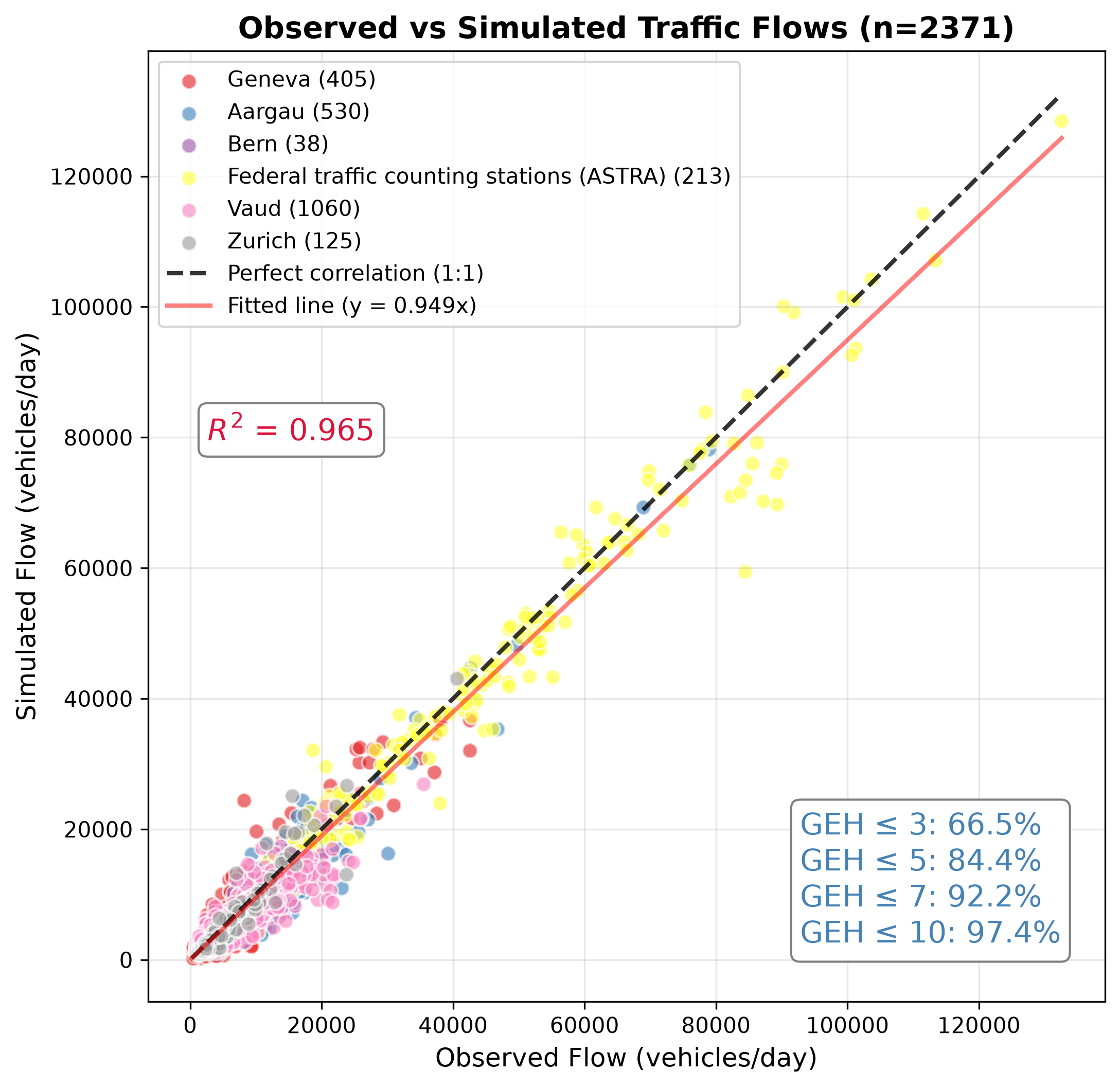}
    \end{minipage}
\end{minipage}
\caption{Observed and simulated weekday traffic at 2,371 federal and cantonal observations. Each point uses the same observed flow in both panels. Colours identify the sources. The dashed line denotes equal observed and simulated daily flows.}
\label{fig:allcounts}
\end{figure}

\begin{figure}[H]\centering
\begin{minipage}{0.8\linewidth}\centering
    \begin{minipage}{0.47\linewidth}\centering
    (a) Without agent calibration\\[2pt]
    \includegraphics[width=\linewidth]{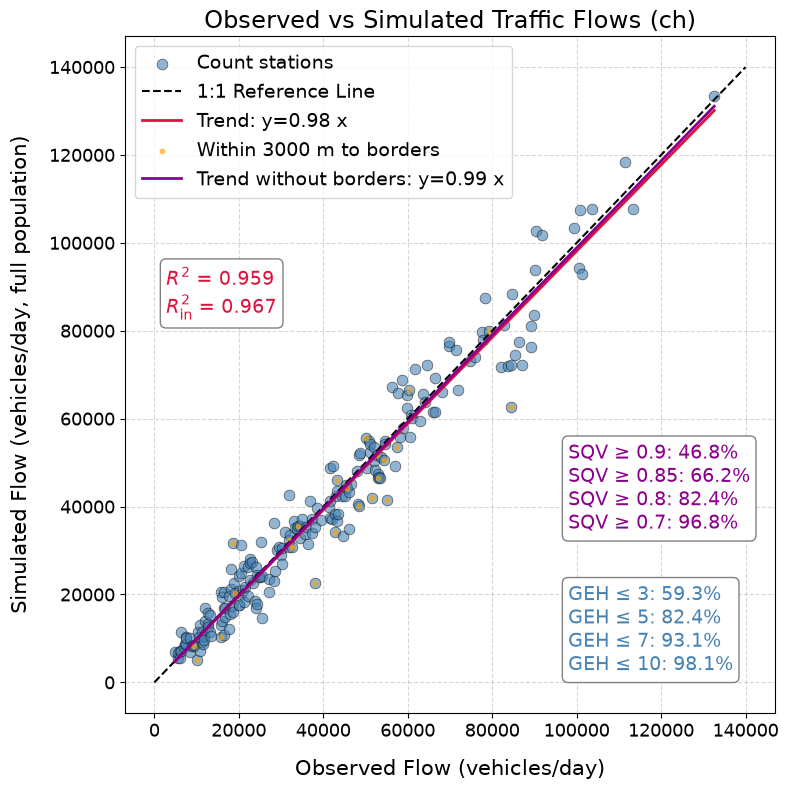}
    \end{minipage}\hfill
    \begin{minipage}{0.47\linewidth}\centering
    (b) With agent calibration\\[2pt]
    \includegraphics[width=\linewidth]{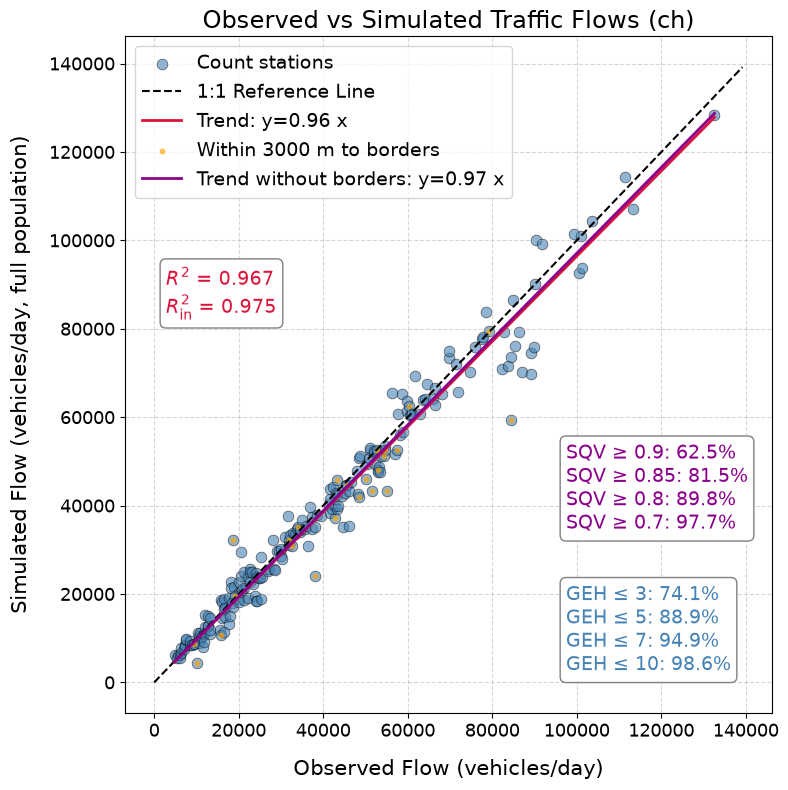}
    \end{minipage}
\end{minipage}
\caption{Observed and simulated daily traffic for the 216 observations in the ASTRA comparison. Orange points identify locations near the border. The statistic $R^2_{\mathrm{in}}$ excludes these locations.}
\label{fig:astra}
\end{figure}

Figures~\ref{fig:allcounts} and~\ref{fig:astra} show observed and simulated daily traffic. Hourly observations were not available for this comparison. To compute the GEH statistic, we divide each daily flow by 24~h in order to convert it into veh/h. Let $O$ and $S$ denote these observed and simulated hourly averages, respectively. Then $\mathrm{GEH}=\sqrt{2(S-O)^2/(S+O)}$. This statistic measures the agreement in average flow and provides no information about the fit during peak hours. For the daily flows $O_i$ and $S_i$, Table~\ref{tab:counts} reports the root mean square error (RMSE) and the coefficient of determination $R^2=1-\sum_i(S_i-O_i)^2/\sum_i(O_i-\bar{O})^2$.

\begin{table}[!htbp]\centering\small
\caption{Agreement with weekday traffic counts at iteration 400. Configuration labels indicate whether individual car ASC calibration was enabled. RMSE is in vehicles per day.}
\label{tab:counts}
\begin{tabular}{llrrrr}\toprule
Observation set & Configuration & $N$ & $R^2$ & RMSE & GEH $\leq5$ (\%)\\\midrule
All available & Without & 2,371 & 0.945 & 3,261 & 73.5\\
 & With & 2,371 & 0.965 & 2,583 & 84.4\\
ASTRA & Without & 216 & 0.959 & 5,348 & 82.4\\
 & With & 216 & 0.967 & 4,770 & 88.9\\\bottomrule
\end{tabular}
\end{table}

Across the 2{,}371 observations, the share with $\mathrm{GEH}\le 3$ rises from 50.7\% to 66.5\%. The share with $\mathrm{GEH}\le 5$ rises by 10.9 percentage points, and the RMSE of daily flows falls by 20.8\%. The ASTRA comparison shows a similar pattern, with a gain of 6.5 percentage points at $\mathrm{GEH}\le 5$ and a 10.8\% reduction in RMSE. This improvement in typical count agreement coexists with a greater underestimation on average: across all observations, the mean signed error changes from $-160$ to $-613$ vehicles per day.

The orange dots, which identify locations close to the Swiss border, show that these points change little between the two simulations. This is because the majority of the simulated flow on these links stems from cross-border traffic, which is treated as fixed background traffic and does not undergo mode choice. Some locations further from the border also have a large share of cross-border traffic and therefore cannot improve further either.

\subsection{Distribution and evolution of the constants}
\begin{figure}[H]\centering
\includegraphics[width=0.9\linewidth]{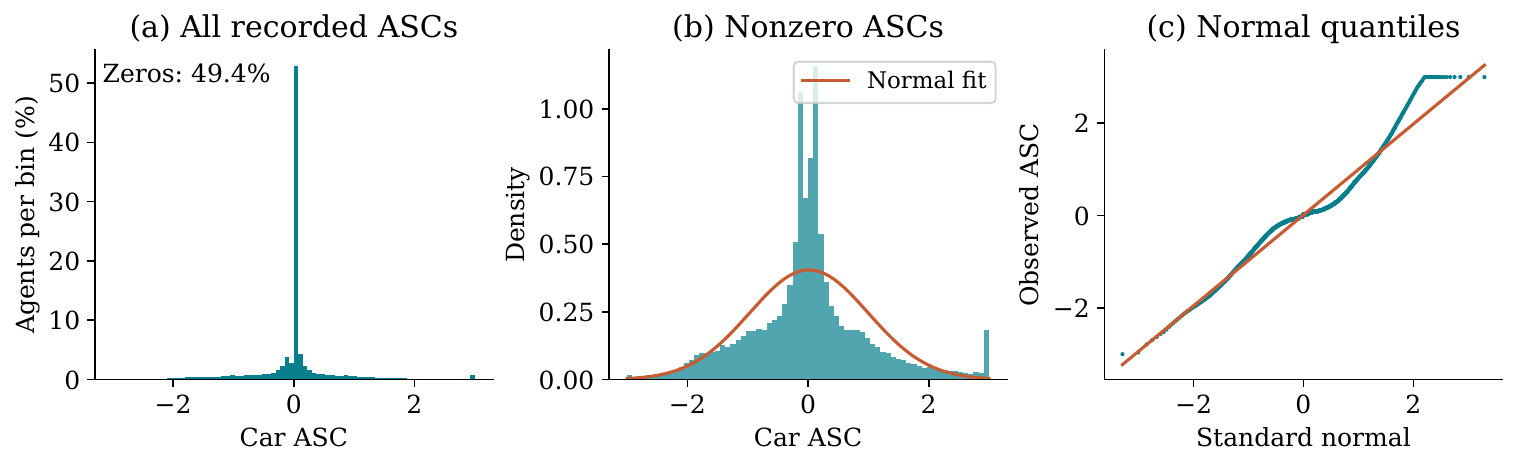}
\caption{Distribution of individual car ASCs at iteration 400. The first panel includes all eligible agents. The middle and right panels compare the nonzero values with a fitted normal distribution. }
\label{fig:distribution}
\end{figure}

Of the 645,589 eligible agents, 319,106 (49.43\%) finish with their initial ASC of zero. Among the remaining 326,483 agents, 164,740 finish with a positive constant and 161,743 with a negative one (Figure~\ref{fig:distribution}). Across all eligible agents the mean is 0.0074 and the standard deviation is 0.700. Among agents with nonzero values the mean is 0.015, the standard deviation is 0.985 and the 5th and 95th percentiles are $-1.667$ and 1.870. The upper bound of $+3$ is reached by 4,532 agents and the lower bound of $-3$ by 383.

The nonzero values form a broadly bell-shaped distribution with a longer right tail than a normal distribution (skewness 0.392, excess kurtosis 1.284). The normal quantile plot shows departures in both tails. A descriptive Shapiro--Wilk check on 5,000 sampled nonzero values gives $W=0.955$. Agents share OD corrections, which limits an independent-sample interpretation of the test. Limited count coverage offers one explanation for the mass at zero. Some agents make trips between zones with no usable count signal, while for others the daily signals can cancel or fall below the update threshold. The final distribution does not distinguish these cases.

\begin{figure}[H]\centering
\includegraphics[width=0.8\linewidth]{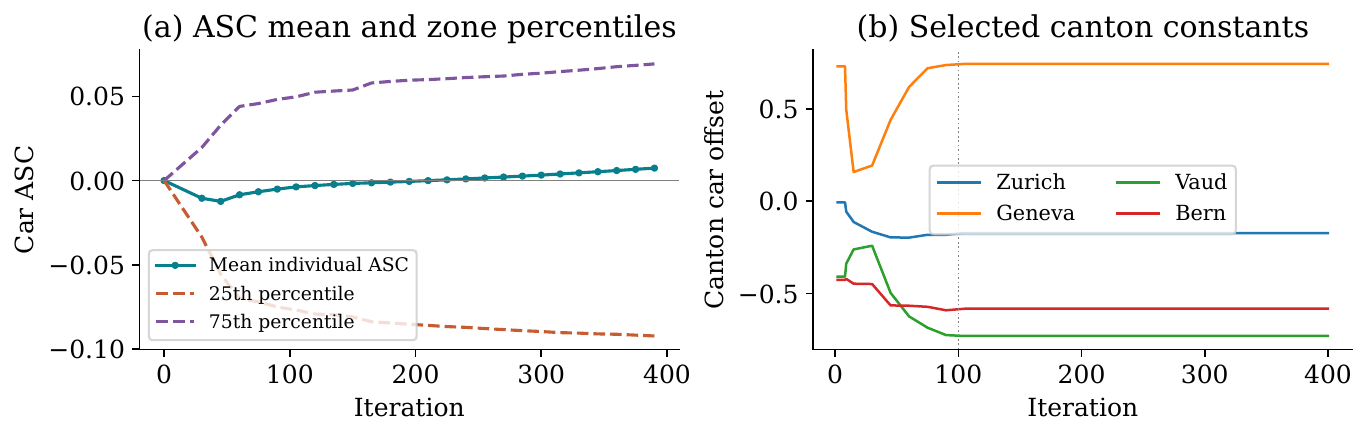}
\caption{Evolution of the calibrated car constants. Panel (a) shows the individual mean and the population-weighted 25th and 75th percentiles of home-zone means. Panel (b) shows selected cantons car ASCs.}
\label{fig:evolution}
\end{figure}

The mean individual ASC stays close to zero as home-zone means spread apart (Figure~\ref{fig:evolution}a). Early changes partly reflect grid refinement. On the final grid, the population-weighted mean absolute change in zone means was 0.00194 between iterations 165 and 180 and still 0.00170 between 375 and 390. The last update therefore continued to shift zone means. This measure does not resolve individual changes within zones or their effect on count fit. Canton offsets (canton level ASC) in Figure~\ref{fig:evolution}b are fixed after iteration 99 by design.

\subsection{Mode shares}
\begin{figure}[H]\centering
\includegraphics[width=0.55\linewidth]{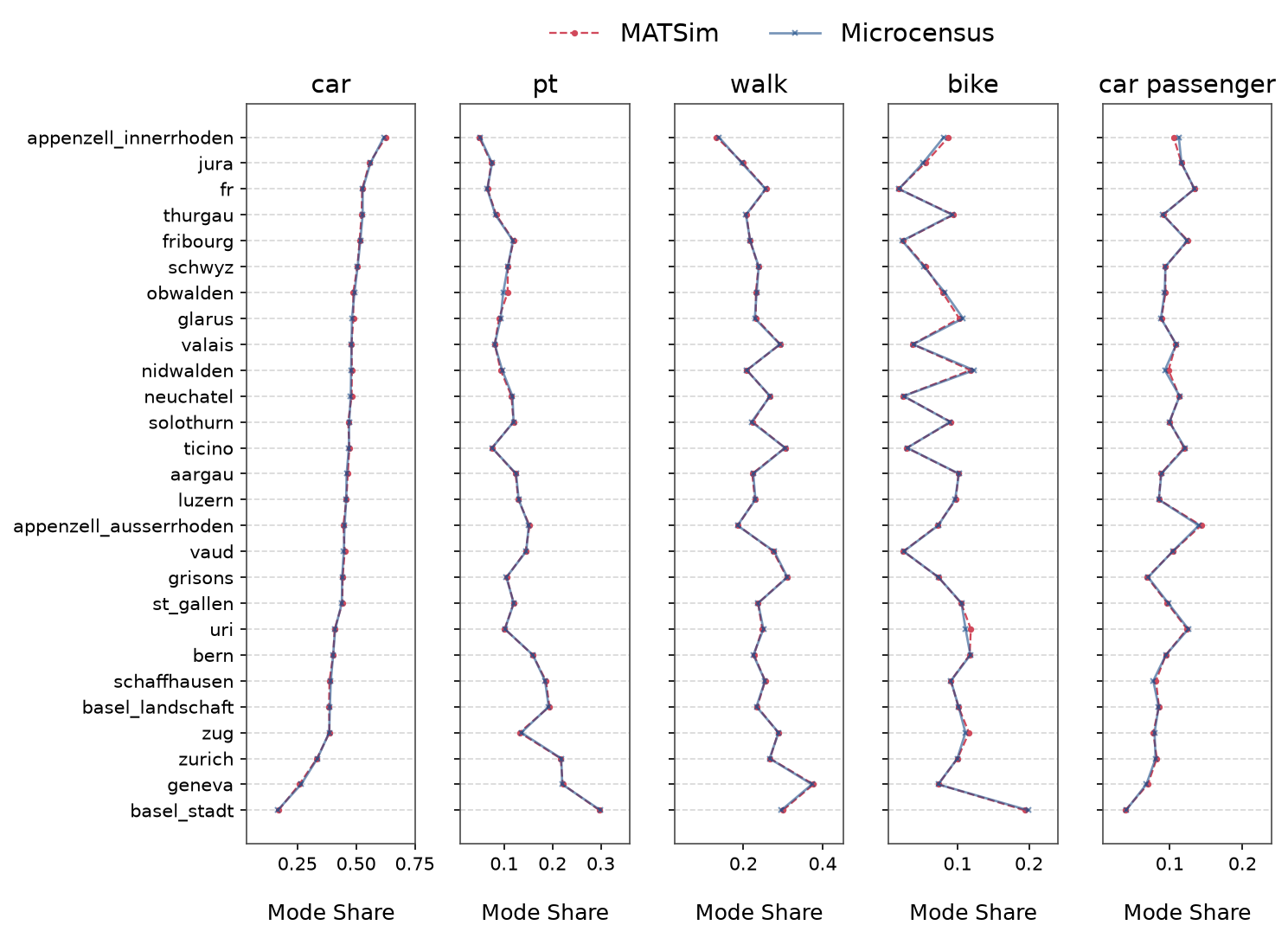}
\caption{Mode shares by canton and mode.}
\label{fig:cantonshares}
\end{figure}

The mode shares of the simulated trips in the simulation with agent-specific ASC calibration closely follow the Microcensus targets for all five simulated modes (Figure~\ref{fig:cantonshares}). This shows that the calibration also ensures that the mode shares are respected for every mode and within every canton. A similar figure was presented in \citep{dib2026online, dib2026mum}. The difference here is that car passenger trips are not fixed and can undergo mode choice like all other modes.

\begin{figure}[H]\centering
\begin{minipage}{\linewidth}\centering
\begin{minipage}{0.485\linewidth}\centering
(a) By trip crow-fly distance\\[2pt]
\includegraphics[width=\linewidth]{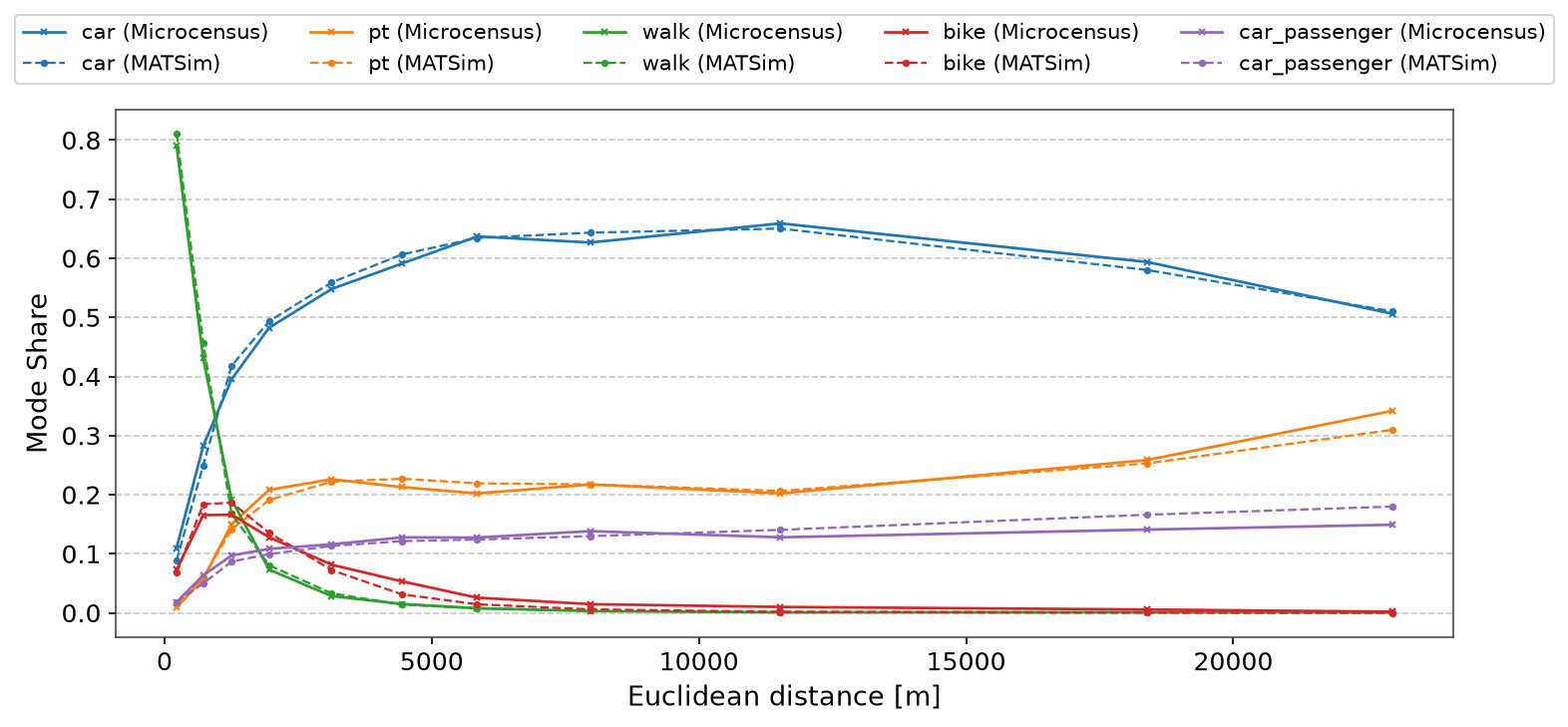}
\end{minipage}\hfill
\begin{minipage}{0.5\linewidth}\centering
(b) By age\\[2pt]
\includegraphics[width=\linewidth]{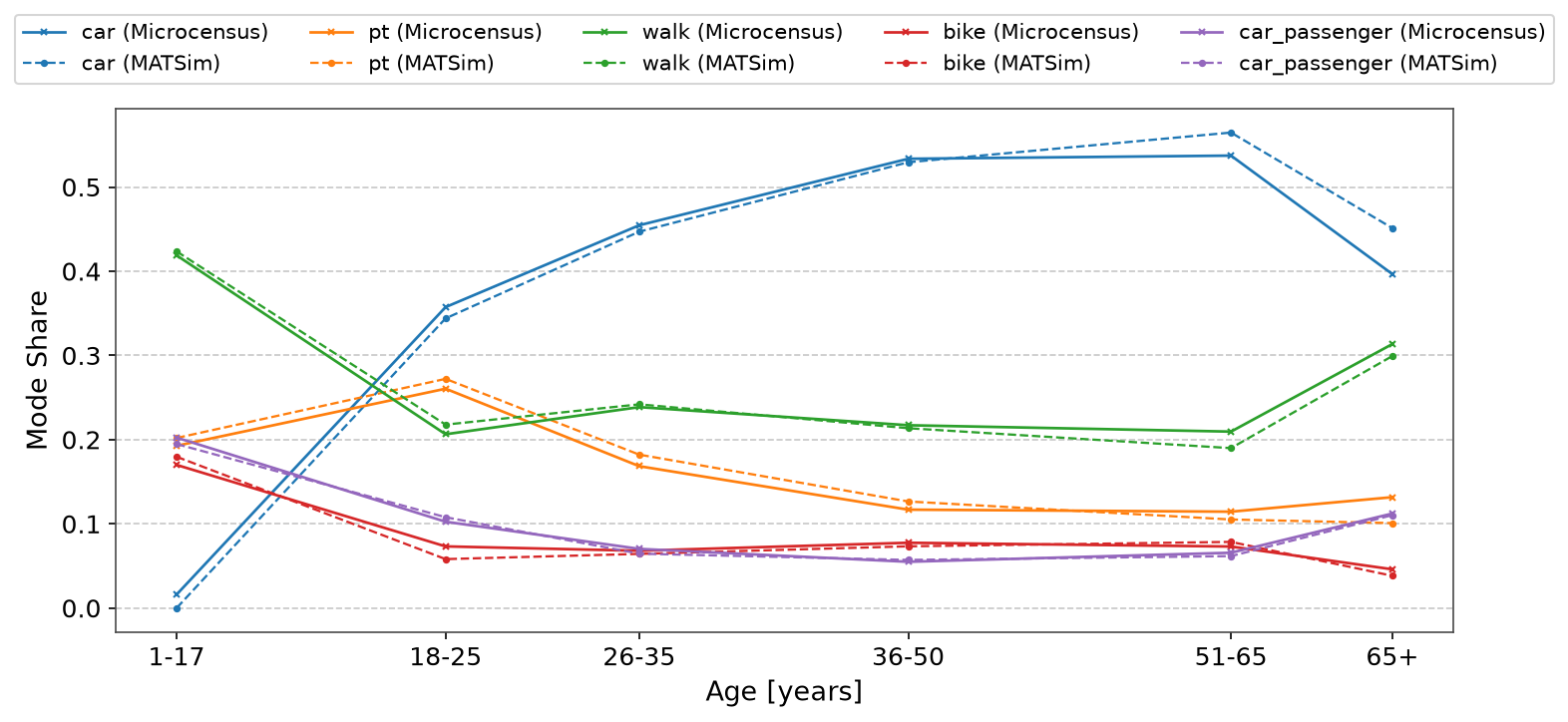}
\end{minipage}
\end{minipage}
\caption{Mode shares by crow-fly trip distance and agent age.}
\label{fig:sharesdistanceage}
\end{figure}

When we examine the mode shares within each trip-distance bin and each age bin (Figure~\ref{fig:sharesdistanceage}), we find that they closely follow the Microcensus. This indicates that agent-specific ASCs do not distort the mode shares within these population segments.

\section{Discussion}
Count agreement improves in the Swiss run with agent-specific car ASCs. Across 2,371 observations, the proportion with GEH at most five rises by 10.9 percentage points over the baseline. Daily-flow RMSE also falls, although the mean signed error becomes more negative. For an activity-based model, the practical result is that a better fit to observed traffic is attainable while keeping the origins and destinations of each agent's trips fixed in the mode-choice update.

Several count-based methods can compensate for errors elsewhere in a demand model. If the synthetic population assigns too many transit subscriptions in a corridor, or overstates household car access, OD calibration can absorb the resulting flow error by changing trip totals between zones, Cadyts can favour plans that better reproduce the counts, and an agent ASC can change the attractiveness of driving on trips for which a car is available. Each mechanism can improve the fit at a station without identifying the underlying error, which requires independent observations of travellers and their available modes.

Other discrepancies warrant changes elsewhere in the model. An incorrectly coded road capacity or route cost can divert traffic from a counted link, and too few activities in an employment centre, or workplaces assigned to the wrong side of a river, can leave too few trips that could plausibly pass a station. Raising the car ASC in these cases may draw passengers or cyclists into cars while leaving the underlying error in place. Independent evidence on activity locations, trip volumes and network travel times is therefore central to deciding when to adjust mode choice and when to revise demand or supply.

Synthetic population generation is a further source of such errors. Different random seeds can produce different local activity patterns even when aggregate statistics agree \citep{dib2026diary}, so ASCs calibrated on one realisation of the population may partly absorb features specific to it. It is worth analysing the calibrated constants locally, for example by OD pair or corridor, to identify which are stable and which reflect the particular activity patterns generated.

The choice of calibration method should follow the evidence available for each decision. Where trip volumes or destinations are uncertain, OD calibration allows them to respond to counts. Where independent data support the activity locations and trip totals, agent ASCs allow counts to inform which of those trips are driven. In MATSim this also provides an alternative to the standard Cadyts plan-score correction for applications that select modes through the single-plan discrete mode choice extension.

\subsection{Convergence, validation and extensions}
The continuing movement in Figure~\ref{fig:evolution}a raises a practical question: at a given OD pair, how much of the remaining count discrepancy can still be reduced by changing the modes of its travellers? Future work will test an OD-specific learning rate based on the number of eligible trips that replan, the stations those trips serve and the change in car use following earlier adjustments. Where several OD pairs serve the same station, the update should account for their combined contribution. Where successive changes yield no improvement beyond ordinary iteration-to-iteration variation, the local update could fall to zero, with monitoring continuing so that the OD pair can respond again if traffic conditions change.

A first validation experiment would repeat the comparison across random seeds, including synthetic populations generated with different seeds \citep{dib2026diary}, and reserve stations on entire corridors from all count-based calibration. This would allow the local stability of the constants to be assessed against variation in activity patterns. A comparison with OD adjustment on the same held-out corridors would test when retaining activity locations is advantageous. The open eqasim scenario for Paris and {\^I}le-de-France provides a setting in which others could reproduce this experiment \citep{hoerl2021population}.

With observed hourly flows, the method could also estimate separate corrections for morning and evening trips, addressing opposite discrepancies on the same corridor separately. The same principle could extend to public transport using boardings or passenger loads and to cycling using bicycle counts, provided trips are linked to what the instrument measures; for instance, a transfer can generate two boardings on a single public transport journey. Previous MATSim work on public transport path choice provides a starting point \citep{moyo2016, preto2025differentiation}. Keeping base-year constants fixed in policy scenarios would allow service or price changes to affect mode choice. More accurate spatial flows could also inform emissions modelling \citep{dib2025emissions,dib2027emissions} or noise exposure assessment \citep{lebescond2023population}, subject to separate validation of local traffic

\section{Conclusions}
Calibrating car ASCs offers an alternative to direct OD adjustment when the geography of activity demand has independent support. In the Swiss application, the run with individual ASCs reproduced a greater share of the observed flows within both GEH thresholds of 3 and 5, while the agents retained all their activity locations and the global and cantonal mode shares remained accurate. Introducing the correction at the point of mode choice makes the approach suitable for single-plan MATSim applications. The continued movement of the constants late in the run motivates the development of dynamic learning rates that respond to the remaining scope for mode shifts in each OD pair.

\section*{Code and data availability}
The \href{https://github.com/eqasim-org/eqasim-switzerland}{Swiss scenario-generation code} and the \href{https://github.com/eqasim-org/eqasim-java/tree/cmdp/core/src/main/java/org/eqasim/core/components/network_calibration/demand_calibration/agent_ascs}{agent ASC calibration implementation} are publicly available. Reproducing the Swiss population requires population and mobility microdata obtained from the Swiss Federal Statistical Office (BFS) under its access conditions.

\small
\setlength{\bibsep}{3.5pt plus 0.5ex}   
\renewcommand{\baselinestretch}{0.9} 
{\color{blue}\bibliography{references}}
\end{document}